\documentclass[10pt,journal,compsoc]{IEEEtran}
\ifCLASSOPTIONcompsoc
  \usepackage[nocompress]{cite}
\else
  \usepackage{cite}
\fi
\ifCLASSINFOpdf
  \usepackage[pdftex]{graphicx}
  \graphicspath{{./}}
  \DeclareGraphicsExtensions{.pdf,.jpeg,.png}
\else
  \usepackage[dvips]{graphicx}
  \graphicspath{{./img/}}
  \DeclareGraphicsExtensions{.eps}
\fi
\usepackage{color}

\usepackage{amsmath}
\usepackage{algpseudocode}

\begin{document}
%
\title{Blockchain based trust \& authentication for decentralized sensor networks}
%
%
%
%

\author{Axel~Moinet,
        Benoit~Darties,
        and~Jean-Luc~Baril,
\IEEEcompsocitemizethanks{\IEEEcompsocthanksitem Axel Moinet, Beno\^it Darties and Jean-Luc Baril are from the Le2i laboratory, FRE CNRS 2005, Univ. Bourgogne Franche Comt\'e.\protect\\
E-mail: axel.moinet@u-bourgogne.fr}%
\thanks{Manuscript received February 1, 2017; revised February 1, 2017.}}

\IEEEtitleabstractindextext{%
\begin{abstract}
Sensor   networks   and   Wireless   Sensor   Networks
(WSN)  are  key  components  for  the  development  of  the  Internet of Things. These networks are subject of two kinds of
constraints. \textit{Adaptability} by the mean of mutability and evolutivity,
and  \textit{constrained  node  resources}  such  as  energy  consumption,
computational complexity or memory usage. In this context,
none of the existing protocols and models allows reliable peer
authentication  and  trust  level  management.  In  the  field  of
virtual  economic  transactions,  Bitcoin  has proposed  a  new
decentralized  and  evolutive  way  to  model  and  acknowledge
trust  and  data  validity  in  a  peer  network  by  the  mean  of  the
blockchain. We propose a new security model and its protocol
based  on  the  blockchain  technology  to  ensure  validity  and
integrity  of  cryptographic  authentication  data  and  associate
peer  trust  level,  from  the  beginning  to  the  end  of  the  sensor
network lifetime.
\end{abstract}

}

\maketitle

\IEEEdisplaynontitleabstractindextext

%
\IEEEpeerreviewmaketitle

\IEEEraisesectionheading{\section{Introduction}\label{sec:introduction}}

%
%
%
%
\IEEEPARstart{S}{ensor} Networks and Wireless Sensor Networks (WSN) are two main components involved in the development of the
Internet of Things (IoT). Security and privacy handling for Sensor Networks present new issues due to specific constraints. Low resources on computation, hardware functionalities and energy consumption in WSNs. We can divide research work into two categories: security and privacy for the data being sent over the network on one side, and node authentication and trust management on the other side. 
Both have been actively explored the last ten years, and some solutions have been brought by researchers. However, from our knowledge none of these works propose a complete model for both content access, security, privacy and trust management. In this paper, we focus on addressing authentication and trust management issues.

\subsection{Overview}

During our researches, we have separated existing work into two distinct research areas. The first one is authentication and trust management issues in decentralized networks and WSN. Then we consider ongoing work on blockchains and their applications.

\subsubsection{Authentication and trust for decentralized networks}
We can find a lot of different approaches for authentication in WSN and the IoT in the literature. As outlined by Medaglia {\textit et al.}\cite{Medaglia2010}, WSNs have specific security constraints on node authentication to ensure data validity and confidentiality.
Trust management is tied to authentication mechanisms, as a the mean to identify the trustee and the truster. We take previous work on trust evaluation in distributed networks by Sun {\textit et al.}\cite{sun2006trust}, as a reference on issues concerning trust in decentralized networks for our work.

\subsubsection{Blockchain as a secured data structure}
Recent work by Zyskind et al.\cite{Zyskind2015} shows the interest of the blockchain as a personal data management platform focused on privacy. They outlined how the blockchain helps leveraging user control over data in the context of social networks and big data. Foutiou et al.\cite{Fotiou2016} describe a decentralized name based security system using blockchains to secure contents access in Information-Centric Networking based architectures. These approaches prove usability of the blockchain as a secure decentralized data structure for new applications, but none has been used to provide node authentication and trust management in Wireless Sensor Networks (WSN) and in the Internet of Things (IoT). 

\subsection{Our Contribution}
We propose a model based on blockchain data structure used to store decentralized authentication and node trust informations. This model is evolutive, adaptative and ensure reliability over time.

\subsection{Organization}
We first explain briefly the blockchain data structure as presented in Bitcoin. Then we present issues in decentralized node authentication and trust management for WSN. The last part of this paper describes our model of a blockchain based solution for authentication and trust management which provide a solution to overcome decentralized networks issues.

\section{The blockchain technology}
In 2008, a person or group of persons known under the name of Satoshi Nakamoto published a paper\cite{Nakamoto2008} dealing with a new decentralized peer-to-peer electronic cash system. This paper introduces the blockchain as a new data structure to store financial transactions, as well as an associate protocol to ensure the validity of the blockchain in the network.

\subsection{Data structure}

In his paper, Nakamoto describes the blockchain as a database modeled by a linear sequence of blocks, each one containing cryptographic hashes corresponding to the previous and current block to ensure continuity and immutability. Bitcoin uses the blockchain to store financial transactions and contracts.

\begin{figure}[h]
\centering
\includegraphics[width=0.5\textwidth]{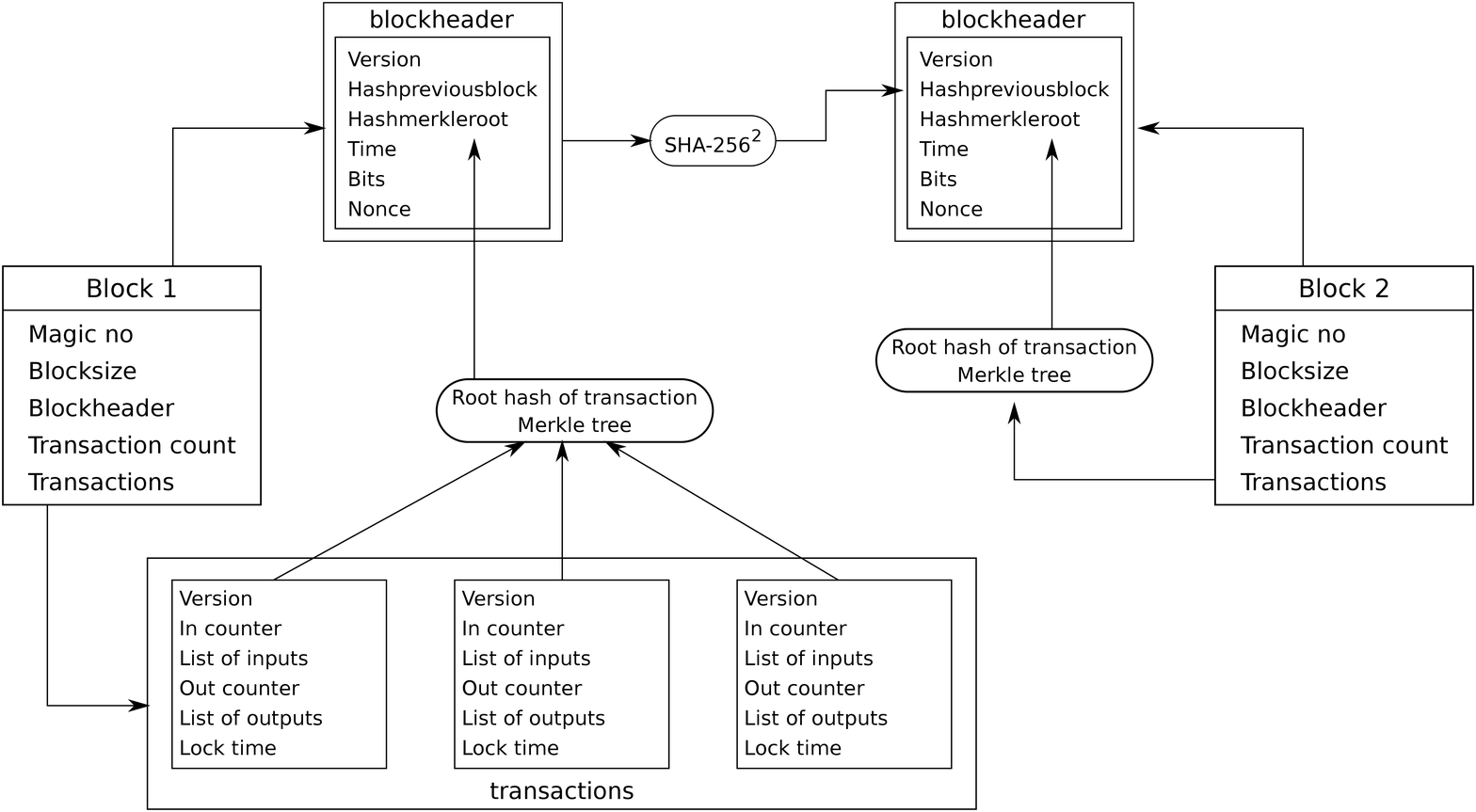}
\caption{Bitcoin block chaining mechanism. The Merkle root of all transactions is included in the block header and then used as input for the next block in the chain.}
\label{fig:Bitcoin-block-chaining}
\end{figure}

The chaining method used in Bitcoin (Figure \ref{fig:Bitcoin-block-chaining}) ensures the immutability by using the hash of the previous header block hash in the current block. The header includes the root hash of the Merkle tree of all transactions in the block. This way transactions cannot be changed without changing the root Merkle hash and then invalidating the block. Due to the way the blockchain is built, fork chains can append with different valid blocks storing different transactions. The Bitcoin protocol resolves this issue by selecting the longest blockchain as the correct one. Note that due to this choice, even after being included in a valid block, transactions can be considered valid only after a subsequent block has been calculated and successfully included in the blockchain by the majority of the network \cite{eyal2016bitcoin}.

\subsection{Secure distributed storage based on blockchains}
\begin{figure}[h]
\centering
\includegraphics[width=0.5\textwidth]{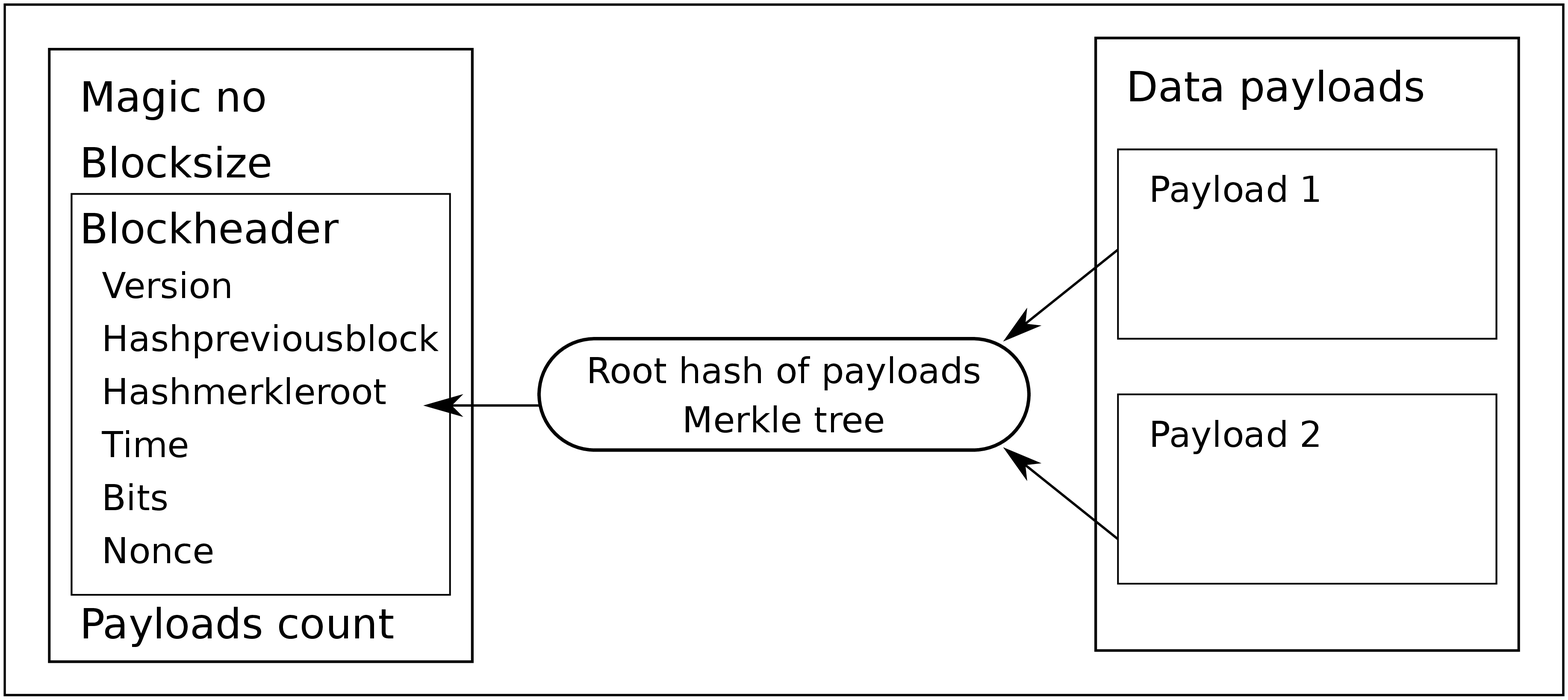}
\caption{Blocks used as generic storage. We use structured payloads and replace transactions by payloads in the Merkle root hash computation.}
\label{fig:generic-block-parts}
\end{figure}

We consider the blockchain data structure outside of its application in Bitcoin, as a generic decentralized secured data storage structure. It is possible to use any data payloads other than transactions as parts of the block. The block is then divided in two parts, (a) the block constants and header and (b) the data payloads, as shown in Figure \ref{fig:generic-block-parts}.

A single modification in one payload of a block will change its Merkle root hash value, and then invalidate it. This solution thus provides secure and reliable storage distributed among all peers in the network. Note that this implies that the complete blockchain and all datas linked to it must be duplicate on all peers. The size of data payloads will influence both the hash calculation algorithm and bandwidth used to maintain the blockchain. Due to bandwidth restriction in the platform we use, we choose to limit the total size of a block to 5 MB to ensure we have enough storage for important security and trust informations without overloading the network with blockchain control data.


\section{Authentication and Trust in WSN}

Authentication and trust can be seen as two parts of the same problem \cite{solhaug2007trust}. Authentication allows us to be sure to who we are dealing with, trust giving us insights of how we can rely on and dealing with a potential risk on an action. If we consider the presence of a master authority in charge of authentication and trust management, we can easily ensure a good security and privacy level in the network. However, this has a major drawback, the master authority becomes the central part of the network security and thus the critical point of vulnerability in the network. In decentralized and ad-hoc networks, this approach is impossible, because we don't have a node which can assure to be connected at every moment of the network life.




\subsection{Our Framework}

To ensure proper organization and content management in decentralized networks, we use a common content model based on Service Oriented Architecture\cite{erl2005service} adapted for our application and compatible with CoAP protocol\cite{shelby2014constrained}. This approach allows us to design a RESTful model for interaction with the internet, and his currently outlined at one promising approach to organize sensor networks\cite{de2011service}.

\subsubsection{Network services model}
Before further introspection on our blockchain based model, we must define the network model we use. Wireless Sensor Networks can be well described as decentralized networks composed of resource constrained nodes based on embedded devices. We choose to model the network as an undirected graph \(G=(V,E)\), each vertex describing a node in the network, and each edge links two nodes within transmission range from each other. Then we associate abilities and services to nodes, providing resources on the network. In this model we define two entities formalized as a set of characteristics vectors.
\begin{itemize}
\item Network Node (NN) defines a vector of Node Properties (NP) and another of Node Abilities (NA)
\end{itemize}
  
\begin{equation}
NN = \begin{pmatrix}
  NP = \begin{bmatrix}
    name & energy & cpu \\
  \end{bmatrix} \\
  NA = \begin{bmatrix}
    camera & storage \\
  \end{bmatrix}
\end{pmatrix}
\end{equation}%

\begin{itemize}
\item Available Services (AS) defines an Abilities Dependencies (AD)  vector, a Resources Dependencies (RD) vector and a Resources Provider (RP) vector
\end{itemize}

\begin{equation}
AS = \begin{pmatrix}
  AD = \begin{bmatrix}
     camera & storage \\
  \end{bmatrix} \\
  RD = \begin{bmatrix}
     \\
  \end{bmatrix} \\
  RP = \begin{bmatrix}
    videostream & videorecording \\
  \end{bmatrix}
\end{pmatrix}
\end{equation}%

Each node stores services in a Service Registry (SR). Nodes having the storage ability can store services they cannot deploy to ensure reuse of these services in the future on other nodes.

In the next section, we refer to our service model and related abbreviations to describe our solution providing authentication and trust management mechanisms for decentralized networks.

\section{Blockchain Authentication and Trust Module (BATM)}

Public Key Infrastructure (PKI) is a major component to resolve authentication in networks. In 1991, Zimmerman introduce a new concept named web of trust for his Pretty Good Privacy (PGP) encryption program~\cite{zimmermann1995official}, which was then standardized by the IETF under the OpenPGP name. Current version of the standard is described in RFC 4880\cite{callas2007rfc}. OpenPGP use PKI to provide three main functionalities.

\begin{itemize}
\item Confidentiality with Encryption
\item Authentication via Digital Signature
\item Web of Trust via identity validation from peers
\end{itemize}

BATM proposes a new way to achieve these goals using the blockchain as the database to store public keys, digital signature and peer informations, allowing each component of the network to validate informations about every other node in the network.

This section explains the global design of BATM in regard to three aspects. First, we focus on authentication, public keys, block mining and their mutual influence. Then we explain principles and particularities of the block exchange protocol and associate rules. Finally, we describe how the combination of authentication and protocol rules allows to define a trust management model.


\subsection{BATM authentication}

BATM associates cryptographic keys with each NN and AS in the network. We use the idea contained in the PGP model of a master key to identify a NN or AS among its lifespan. This key is only used to generate secondary keys for encryption and digital signature. As in most PKI, private keys are the main component of the system, and so key management is particularly critical. An attacker can easily spoof NN identity if he retrieves its keys. In this regard, implementations will need to be careful in the choice of the keyring to store private keys, but we won't address this issue in this paper.

\subsubsection{BATM block mining}
We assimilate each data payload as an event providing informations about the status of a NN and its cryptographic informations. At authentication, a node submits a credential payload containing its master public key along with secondary keys. We ensure key renewal to mitigate attacks known for guessing keys by using key validity timeouts.

As network security and privacy relies on informations contained in the blockchain, our design forbid to add blocks uniquely by resolving the problem and satisfy header hash requirement. More precisely, only authenticated nodes can mine new blocks, and only if they haven't issued a payload to be included in the block. To fullfill these requirements, miners must choose which payloads to include in the block they try to resolve.

To be valid, a block must both resolves the problem and contains a valid Miner Approval (MA) payload generated by the Miner, illustrated by the algorithm in Figure \ref{alg:block-validation}. This kind of payload contain a digital signature of a random value contained in the previous block MA payload, and must correspond to a successfully authenticated node.

\begin{figure}
  \begin{algorithmic}[1]
    \Require currentblock previousblock
    \Ensure block validity
    \If{ not(HashCurrentBlock resolves problem)}
        \State return false
    \EndIf
    \If{ not(MinerApproval payload valid)}
        \State return false
    \EndIf
    \If{ CurrentBlock has event payload for miner NN}
        \State return false
    \EndIf
    \If{ not(all payloads in block valid)}
        \State return false
    \EndIf
    \State return true
  \end{algorithmic}
  \caption{Block validity check algorithm.}
  \label{alg:block-validation}
\end{figure}

\subsubsection{BATM data payloads}

When a NN or AS requests to join the network for the first time, it issues a specific Credential Payload (CP) to all NNs. A CP contains public keys needed to operate in the network. Authentication request is approved when an authenticated NN includes the CP in a valid block.

Credential status of the NN / AS can be subsequently updated by renew payload and revoke payload. Note that when revoking his credential, a NN / AS must provide a new credential payload to remain authenticated in the network. Miners will try to include revoke payload and new credential payload in the same block to ensure continuity of node status in the network.

If we allow submission of payloads without further verification, every node could be allowed to propose payloads in the network. To overcome this issue, payloads use a system of signed hash digests. Every payload must have a hash digest signed by payload issuer as its last entry. This way, our payload verification algorithm can easily check the validity of the data. Note that revoke payload use the master key to sign data hash, whereas other payloads use the current signature subkey.

BATM uses 6 different payload types as follows.
\begin{itemize}
  \item MA (Miner Approval)
  \item NN and AS Payloads
  \begin{itemize}
    \item Credentials
    \item Renew
    \item Blame
    \item Ban
    \item Revoke
  \end{itemize}
\end{itemize}

\begin{figure}
\centering
\includegraphics[width=0.45\textwidth]{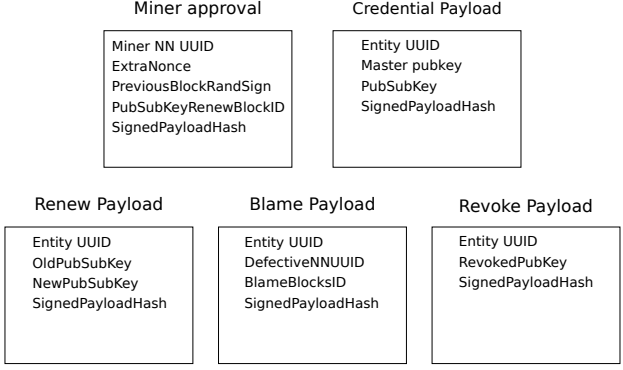}
\caption{Data payloads available in BATM. All blocks must have a Miner Approval payload to be valid, to verify which node has allowed the authentication entry.}
\label{fig:batm-data-payload}
\end{figure}

We provide a detailed description of data contained in BATM payloads in Figure \ref{fig:batm-data-payload}. Note that Blame and Ban payloads are specific payloads used in BATM trust management model.

We showed how our model of a PKI using the blockchain ensure reliable storage for cryptographic material, and how we use it to perform NN authentication on the network. We then propose a trust management model using informations contained in the blockchain.

\subsection{BATM trust management}


The first need for a definition of trust originated from social studies to characterize relations between people in the society. In this context, we consider relevant to use Gambetta's definition of trust\cite{Gambetta1990} as an assumption on the level of subjective probability about how a particular agent will perform an action from a subjective point of view. Note that we understand the term of subjective probability as a reputation level applied to the realization correctness of a subsequent action in the future.

This interpretation of trust implies that the reputation level associated with an agent must vary over time to match the actual realization of the action. Good behaviour must be rewarded, and bad behaviour must be punished to maintain accurate prediction on the realization of actions.

\subsubsection{Knowledge based trust for BATM}

The BATM module includes a trust model called Human-like Knowledge based Trust (HKT), based on human like behaviour to maintain a reputation level for each node. HKT is a compromise between a mutual surveillance by all nodes on the network and the presence of a trust center.

We use the payloads contained in the blockchain as an indication of each node behaviour on the network over time. This way, we ensure a node cannot fool others by tampering data or pretending to be someone else. Thus we ensure reliability of trust evaluation without the need of a trust center. Following development will be targeted at NN trust evaluation, but same principles apply to AS, with the particularity that AS reputation level is echoed on each node in the network, thus modifying reputation level on each node using it.\\

For each payload type, HKT defines events and associates them reputation factors. We note \(C_{evt}\) the reputation factor for the event, and \(T_{evt}\) the time the event occured.

To make the NN reputation evolve naturally over time, each event reputation factor must be weighted by a function evolving in time since the event occurs. As we want to decreasing contribution of a particular event to the NN reputation level over time, we need to use a continuous decreasing function such as \(e^{-x}\).

During it first authentication, a NN has no passed action to compute a reliable trust value. Thus we choose to grant a base trust value to all nodes when a trusted node gives them access to the network by including their credentials in the blockchain.

For the simulations, we used the following values for event reputation factors.

\begin{itemize}
\item \(C_{approval} = 1\)
\item \(C_{auth} = 8\)
\item \(C_{renew} = 2\)
\item \(C_{blame} = -8\)
\item \(C_{ban} = -16\)
\end{itemize}

We have estimated the following formulas to determine the reputation of node over time.\\

  \begin{equation}
    \forall evt \in (N,Blk(t)): C_{N,t} = \sum C_{evt}
  \end{equation}
\vspace{-.2cm}
  \begin{equation}
    Reputation(N,t_{now}) = C_{auth} + \sum_{t=t_{first}}^{t=t_{now}} C_{N,t} *  e^{\frac{-(t_{now}-t)}{256}}
  \end{equation}%

In this formula, \(t_{first}\) corresponds to the first block in the blockchain after node has authenticated. \(C_{t}\) is used as the global coefficient for all events concerning the node at \(t\) (the sum of all \(C_{evt}\) at \(t\)).

\begin{figure}
\centering
\includegraphics[width=0.45\textwidth]{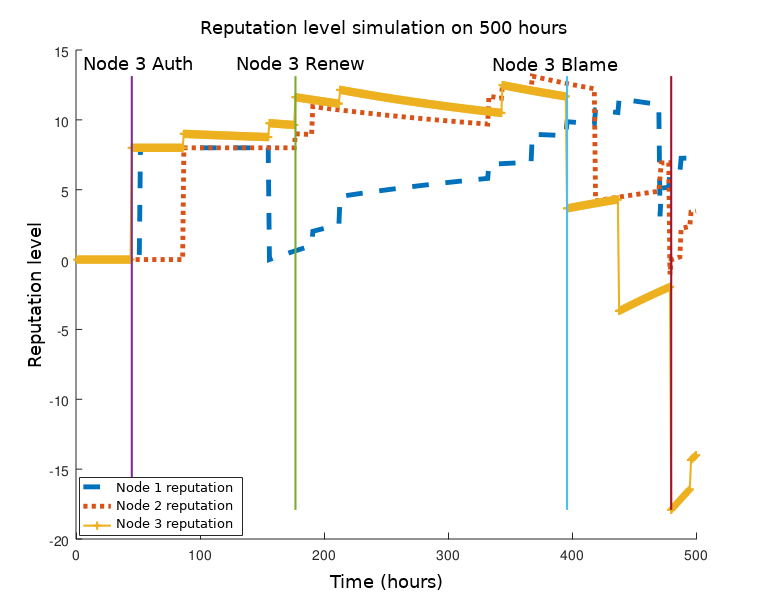}
\caption{Reputation level simulation over 500 hours with 3 NN. After 2 blames, a ban is declared on the node and becomes effective for a 84 hours.}
\label{fig:trust-simul-3-nodes}
\end{figure}

\subsubsection{Trust evaluation}

As we defined it earlier, we consider trust as a probability level that an action will be performed correctly by a NN. In this context, we perform trust evaluation by comparing the current reputation level of a NN to trust him doing certain actions in the network. Currently, we only defined blockchain related actions and associate them with a trust level \(A_{evt}\). This level quantifies the minimum reputation level for a node to be trusted to fullfill subsequent event \(evt\). The trust level is weighted by the number of authenticated NN noted \(N_{auth}\), in order to be less restrictive on actions in a small network and more on huge networks. This behaviour allows NN on the network to work properly and becoming trust defaultive in small network, and then raise the trust level required when more nodes are available.

BATM defines the following trust levels, as equivalent for trust events \(C_{evt}\) described earlier.

\begin{itemize}
\item \(A_{auth}\)
\item \(A_{ban}\)
\item \(A_{blame}\)
\item \(A_{approval}\)
\end{itemize}

We provide following formula to compute values of \(A_{act}\) over time, with \(A_{app}\) being an application factor, allowing applications to be more or less restrictive on actions.

\begin{equation}
  A_{evt} = C_{evt} + \frac{A_{app} * (N_{auth} - 1)}{ C_{evt}}
\end{equation}

Note that reputation level simulation in Figure \ref{fig:trust-simul-3-nodes} does not consider minimum trust level required for a NN to fullfill an action.

We showed how BATM with HKT provide a powerful solution to authenticate nodes and evaluate trust in decentralized networks. The system can be made instable by malicious authenticated nodes overloading the network and submitting lots of valid payloads for inclusion in the blockchain. To overcome this issue, we define specific rules for payload submission in the network to improve stability of the system over time. 

\subsection{BATM payload rules}

To avoid abuse from NN which can overload the network with payloads to be validated, we introduce specific rules on the payload exchange protocol for BATM. We consider two type of rules : timers, key validity timeouts and event reputation factors described earlier. Timers are limitation in time used to discard payloads and blocks submitted by NN overloading the network.

A set of timers defines the minimum amount of time expected between two payloads of the same type. BATM currently uses 3 timers as follows.

\begin{itemize}
\item \(T_{renew}\) is the usual time between two key renewal by a node. If needed, a node is authorized to renew its keys at \(T_{renew} / 2\). Here we ensure there are at most 2 renews in a \(T_{renew}\) for a given NN.
\item \(T_{blame}\) is the minimum time between two blames on a NN given by the same blamer NN.
\item \(T_{banrecover}\) is the time during which a banned NN will be forbidden to mine subsequent block as punishment for a bad behaviour in the network.
\end{itemize}

These timers imply that key validity timeouts must respect the following rules for BATM to work properly.

\begin{itemize}
  \item \(T_{subkey}\) is the timeout for subkeys before renewal. It must be greater than \(T_{renew}\), but less than \(50 * T_{renew}\) to be overcome issue that an attacker may be able to guess the key from data collected in the network.
  \item \(T_{masterkey}\) is the timeout for the master key. In our model, it should be greater than \(10 * T_{subkey}\) and no more than \(50 * T_{subkey}\) to protect it from key guessing attacks.
\end{itemize}

As indication, simulation results showed in Figure \ref{fig:trust-simul-3-nodes} used following timer values, in hours.

\begin{itemize}
\item \(T_{renew} = 168\)
\item \(T_{blame} = 42\)
\item \(T_{banrecover} = 84\)
\end{itemize}

As these rules can be defined to different values regarding the application using BATM, we use the first block in the blockchain to store values to be used. Thresholds will be defined in the future to overcome a problem with a malicious initial NN, and what we called the origin block problem.

\subsubsection{Origin block problem}

At startup, the network contains no authenticated node to realize BATM authentication and trust evaluation, and the blockchain is empty. This means we need a method to forge the first block. We choose to let any node craft this special block from its own parameters. In fact, the main problem is to ensure proper operation in the beginning of the network life, then BATM will adapt itself to events occuring in the network. If the first NN is malicious, it will be banned by others node early in the network.

Since the first block contains all mutable values used in BATM, a possible attack will be the inclusion of specific values which will tend the system to misbehave. To counter this threat, future work will provide a formula to estimate the probability of BATM instability from these values.




\section{Future work}

We presented the concept and model of BATM, with early results on reputation evaluation over time. The next step is to evaluate each part of BATM completely and the global model. The model itself will be improved depending on the results, with the objective of more adaptative algorithms taking AS and NN characteristics as defined in our model.

\subsection{Trust model}

BATM with HKT provide a simple way to manage trust in decentralized networks. More researches on HKT performance must be conducted, and the model itself may evolve to consider more parameters in trust and reputation evaluation. We think about considering NN and AS abilities in account for specific actions, and enhance the reputation calculations. For example, a blamer reputation could influence the reputation factor of its blame, and we may introduce a time coefficient to raise trust on the overall time presence of the NN or AS in the network. Another possibility is to raise resilience over DoS attacks by requiring blames from different NN or AS before banning one.

In this paper, we consider self-organizing networks with no constraints on which NN may ask to authenticate. We also let possible a derivative model, using a network master key to allow blockchain supervision and eliminate the first block problem. A possibility may be to allow specific network, for example vendor-specific networks, where NN can provide a proof of membership by prior network master key signature.

\subsection{Real world testing}

If simulation results fullfill our expectations, BATM will be included in Multicast Services for Linux (MSL), an implementation our SOA network model. Note that MSL is at development stage for now with now release date. Moreover, MSL is intended to be used as a real world proof of concept for our overall design including SOA model and BATM with HKT.

\section{Conclusion}
This paper proposes a new application for the blockchain as a secured decentralized storage for cryptographic keys as well as trust informations in the context of autonomous Wireless Sensor Networks. The Blockchain Authentication and Trust Module and its Human-like Knowledge based Trust model shows how to use to immutability of the blockchain to provide solutions to high problematics in the field of decentralized ad-hoc networks. More precisely, we show how it is possible to build a complete solution providing authentication mechanisms as well as trust evaluation in a self-organized and evolutive network.

\subsubsection*{Resources}

The Service Oriented model is currently under development into the Multicast Services for Linux (MSL) framework. MSL is a free software and will be publicly available at https://bullekeup.github.io/MSL, under the AGPL license. BATM will be available in MSL in the next months. HKT simulation files for GNU Octave and MATLAB are available by mail on request at axel.moinet@u-bourgogne.fr.


%



\ifCLASSOPTIONcompsoc

  \section*{Acknowledgments}
\else
  \section*{Acknowledgment}
\fi

The authors would like to thank the Burgundy Region and the FEDER european fund for funding this research work.

\ifCLASSOPTIONcaptionsoff
  \newpage
\fi



\enlargethispage{.5cm}
\bibliographystyle{IEEEtran}
\bibliography{./blockchains.bib}
%
%
%

%

\begin{IEEEbiographynophoto}{Axel Moinet}
is a Ph.D Student at the Le2i laboratory, Univ. Bourgogne Franche Comt\'e. He works on smart camera based Vision Sensor Networks and combinatorics. He graduated from a Master of Engineering on Embedded Systems and then worked for Oberthur Cash Protection as Embedded Software engineer before returning on a Ph.D thesis. 
\end{IEEEbiographynophoto}

\begin{IEEEbiographynophoto}{Benoit Darties}
is an Associate Professor at the Le2i laboratory, Univ. Bourgogne Franche Comt\'e. He works on combinatorics and operational research applied to networks operations. Benoit obtained in 2007 a Ph.D in Computer Science at University from Montpellier, France. He also works on the design of low energy consumption mac protocols for WSN. 
\end{IEEEbiographynophoto}


\begin{IEEEbiographynophoto}{Jean-Luc Baril}
Jean-Luc Baril received his Phd degree in mathematics and computer sciences from the University of Bordeaux in 1996. He is currently Professor in the Computer Science Department of the Bourgogne University (LE2I), Dijon-France. His research interests are in the areas of combinatorics, graph theory and algorithmic. 
\end{IEEEbiographynophoto}




\end{document}